# Nonperturbative decay dynamics in metamaterial waveguides

*Iñigo Liberal[1,2]\* and Richard W. Ziolkowski[3]*

[1]*Electrical and Electronic Engineering Department, Universidad Pública de Navarra, Campus Arrosadía, Pamplona, 31006 Spain*

[2]*Institute of Smart Cities, Universidad Pública de Navarra, Campus Arrosadía, Pamplona, 31006 Spain*

[3]*Global Big Data Technologies Centre, University of Technology Sydney, Ultimo NSW 2007, Australia*

[*inigo.liberal@unavarra.es](*inigo.liberal@unavarra.es)

**Abstract**- In this work we investigate the nonperturbative decay dynamics of a quantum emitter coupled to a composite right/left handed transmission line (CRLH-TL). Our theory captures the contributions from the different spectral features of the waveguide, providing an accurate prediction beyond the weak coupling regime, and illustrating the multiple possibilities offered by the nontrivial dispersion of metamaterial waveguides. We show that the waveguide is characterized by a band-gap with two asymmetric edges: (i) a mu-near-zero (MNZ) band edge, where spontaneous emission is inhibited and an unstable pole is smoothly transformed into a bound state, and (ii) an epsilon-near-zero (ENZ) band edge, where the decay rate diverges and unstable and real (bound state) poles coexist. In both cases, branch cut singularities contribute with fractional decay dynamics whose nature depend on the properties of the band-edges.

**Main text -** Metamaterial waveguides have been shown to be flexible design platforms for dispersion engineering[1,2]. Salient spectral features in metamaterial waveguides include propagating bands with a negative refractive index, the possibility of opening band-gaps with a prescribed bandwidth, slow-light frequency points with a vanishing group velocity, and anomalous dispersion in highly absorptive bands. Of particular interest are also those frequency points where the propagation constant of the waveguide crosses zero, in direct connection with the field of near-zero-index (NZI) media[3–5]. Points with a near-zero refractive index empower unique wave effects. Examples include perfect transmission through deformed waveguides[6]; intrinsic inhibition of optical turbulence[7]; photonic doping[8], enhanced nonlinear[9], modulation[10] and switching[11] capabilities; and highly directive emission[12,13] to name a few.

Metamaterial waveguides also have wide variety of dispersion profiles. Consequently, they pose an interesting playground for exploring nontrivial quantum light emission. For instance, NZI waveguides can either enhance or suppress spontaneous emission[14–16]. Moreover, the fact that the wavelength is effectively stretched in them enables the observation of superradiance in electrically large samples[17–19], as well as having long-range coherent interactions for entanglement generation and many body physics[20–24]. Recent experiments have shown the enhancement[25], tunability[26], and position independence[27] properties on quantum light emission in NZI metallic waveguides. Furthermore, the experimental demonstration of NZI waveguides in photonic integrated circuits opens the scope and applicability of these concepts even more [28,29].

Most works addressing quantum light emission in NZI waveguides operate within the weak coupling regime or Markovian approximation. Thus, they neglect the dispersion of the waveguide near the frequency of emission. However, a complex dispersion profile is precisely what makes metamaterial waveguides unique. Therefore, the Markovian approximation is found to be too simple and may in fact hide some of the most interesting aspects provided by this class of waveguides. Furthermore, recent theoretical works have highlighted the importance and opportunities offered by taking into account the nonperturbative decay dynamics of quantum emitters coupled to structured reservoirs[30,31], Dirac cone baths[32] and photonic Weyl environments[33].

In this letter we study the nonperturbative decay dynamics for a quantum emitter coupled to a metamaterial waveguide. To this end, we consider the archetypical case of a composite right/left-handed (CRLH) transmission line (TL). This case study is of particular interest as it contains a number of the typical spectral features including a propagating band with a negative index, a band-gap with asymmetric edges, and two frequency points with a near-zero refractive index. The general theory that we are presenting finds applicability in NZI waveguides demonstrated at optical frequencies[25–29], as well as in superconducting circuits[34–36] with equivalent structure.

As schematically depicted in Fig. 1a, we study the coupling of a quantum emitter to a photonic waveguide. This configuration can be modeled with the usual Hamiltonian for a two-level system, $\{e\rangle, |g\rangle\}$, with dipole moment **p** and transition frequency $\omega_0$, coupled to a bath of discrete photonic modes propagating in the waveguide with frequencies $\omega_k$ ($\hbar=1$):

$$H = \omega_0 \sigma^\dagger \sigma + \sum_k \omega_k a_k^\dagger a_k + \sum_k (g_k \sigma^\dagger a_k + h.c.) \quad (1)$$

We focus on the composite right/left-handed (CRLH) transmission line (TL)[1] that can be modeled with the equivalent circuit unit-cell depicted in Fig. 1a. Alternatively, the same system can be described as a one-dimensional (1D) medium described by Drude models with the effective relative parameters $\varepsilon_r(\omega) = 1 - \omega_{ENZ}^2/\omega^2$ and $\mu_r(\omega) = 1 - \omega_{MNZ}^2/\omega^2$ that have epsilon-near-zero (ENZ) and mu-near-zero (MNZ) frequencies corresponding to the electric and magnetic plasma frequencies: $\omega_{ENZ}$ and $\omega_{MNZ}$, respectively[1].

The dispersion relation is given by $k(\omega) = \omega\sqrt{(1-\omega_{MNZ}^2/\omega^2)(1-\omega_{ENZ}^2/\omega^2)}/c$ and its values are presented in Fig. 1b. They tell us that the waveguide is characterized by a low-frequency propagating band with a negative refractive index, a high-frequency propagating band with a positive refractive index, and a band-gap (evanescent wave region) between the MNZ and ENZ frequencies. The refractive index of the waveguide approaches zero at both of these band edges. The coupling strength is given by[14] $g_k = -i\,(\mathbf{p}\cdot\boldsymbol{e}_k)\sqrt{\frac{\omega_k}{2\varepsilon_0 L}Z(\omega_k)\frac{v_g(\omega_k)}{c}}$, where $\boldsymbol{e}_k$ is the unit polarization vector of the modes, $Z(\omega_k) = \sqrt{(1-\omega_{MNZ}^2/\omega^2)/(1-\omega_{ENZ}^2/\omega^2)}$ is the normalized impedance of the waveguide, $v_g(\omega_k) = d\omega/dk$ is the group velocity, and $L$ is the quantization length. In this manner, the waveguide features two points where the refractive index approaches zero, i.e., the MNZ and ENZ frequencies, but at which extremely different impedance and coupling strength behavior is exhibited.

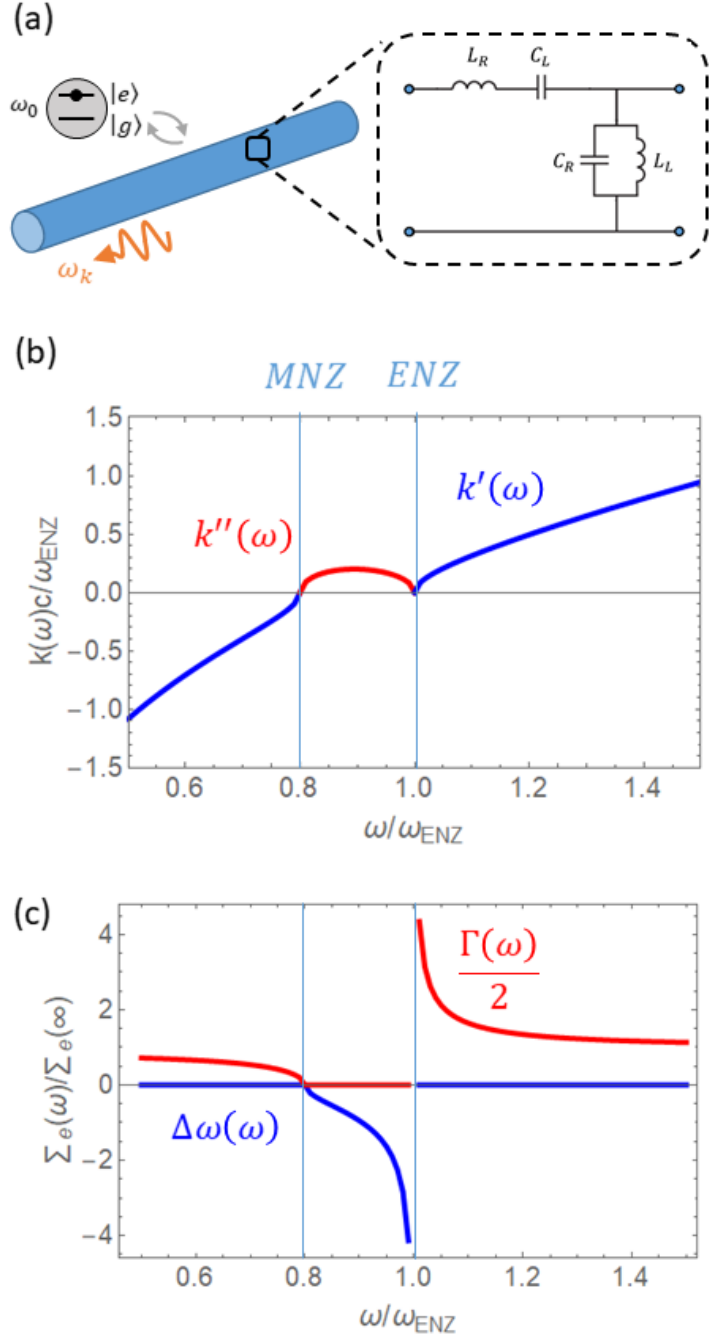

Fig. 1. (a) Sketch of a quantum emitter modeled as a two-level system, $\{e\rangle, |g\rangle\}$, with transition frequency $\omega_0$ that is coupled to a metamaterial waveguide with the equivalent circuit model shown in the inset. (b) Dispersion diagram of the metamaterial waveguide characterized by a band-gap between the mu-near-zero $\omega_{MNZ}$ and epsilon-near-zero $\omega_{ENZ}$ frequencies. (c) Projection of the self-energy on the real axis: $\Sigma_e(\omega + i0^+) = \Delta\omega(\omega) - i\Gamma(\omega)/2$. Its real and imaginary parts describe the frequency shift $\Delta\omega(\omega)$ and decay rate $\Gamma(\omega)$, respectively.

We assume that the system is initially excited, $|\psi(t=0)\rangle = \sigma^\dagger|\{0\}\rangle$, and that it decays into a general single-excitation state $|\psi(t)\rangle = \left[C_e(t)\sigma^\dagger + C_k(t)a_k^\dagger\right]|\{0\}\rangle$. Within this framework, the exact time evolution of the probability amplitudes can be determined by using the resolvent

operator method[37]. Equivalently, one can solve the Schrödinger equation by using the Laplace transform method. By making the change of variables $s = -iz$, the inverse Laplace transform for the probability amplitude of the emitter being excited can be written as an integral over a contour on top of the real axis (see Fig. 2):

$$C_e(t) = -\frac{1}{2\pi i} \int_{i0^+-\infty}^{i0^++\infty} dz \, G_e(z) e^{-izt} \tag{2}$$

with the resolvent

$$G_e(z) = \frac{1}{z - \omega_0 - \Sigma_e(z)} \tag{3}$$

and the self-energy

$$\Sigma_e(z) = \sum_k \frac{|g_k|^2}{z - \omega_k} \tag{4}$$

Evaluating the self-energy in the continuum limit, $\sum_k \to \frac{L}{2\pi} \int dk$, integrating over the frequencies with the replacement $\int dk = \int d\omega \, 1/v_g(\omega)$, applying the Kramers-Kronig relations, and removing the high-frequency divergence leads to the following compact expression:

$$\Sigma_e(z) = -iA \sqrt{\frac{z^2 - \omega_{MNZ}^2}{z^2 - \omega_{ENZ}^2}} \tag{5}$$

where $A = |\mathbf{p}|^2 \omega_0/(4c\varepsilon_0)$. We used a value of $A = 0.01\omega_p$ for the numerical examples so that nonperturbative effects are evident, though the effects are qualitatively the same for other values of $A$ as long as we rotating wave approximation in our Hamiltonian (1) remains valid. The real and imaginary parts of the projection of the self-energy on the real frequency axis, $\Sigma_e(\omega + i0^+) = \Delta\omega(\omega) - i\Gamma(\omega)/2$, correspond to the frequency shift and decay rate, respectively. They are depicted in Fig. 1c. The main characteristic that differentiates the CRLH transmission line from other dispersive and slow-light waveguides is the presence of its asymmetric band-edges. Although the group velocity becomes zero at both band-edges, the decay rate diverges at the ENZ frequency and vanishes at the MNZ frequency in accordance with a previous study in the weak coupling regime[14].

The dispersion properties of the self-energy exemplify the great degree of design flexibility offered by metamaterials waveguides, particularly near the band-edges. On the one hand, the divergence of the decay rate near the ENZ frequency enables the enhancement of the emitters decay rate (brightness), as well as a more efficient (deterministic) coupling to the waveguide mode. The soft transition at the MNZ frequency enables tuning between the dissipative and collective interactions. Moreover, the fact that these responses take place at points where the propagation constant vanishes ($k = 0$) suggests that our configuration is an interesting platform for investigating long-range collective interactions, which are of interest for entanglement generation and many-body physics.

By using complex analysis techniques, the integration can be closed in the lower half-plane; and the time evolution of the probability amplitude can be rewritten with the residue theorem as the sum of the contributions from the different singularities in the complex plane. As schematically depicted in Fig. 2, they consist of three poles and two branch cuts. Thus, the probability amplitude is given by their contributions as

$$C_e(t) = R_1 e^{-iz_1 t} + R_2 e^{-iz_2 t} + R_{BS} e^{-ix_{BS} t} + C_{BC\_ENZ}(t) + C_{BC\_MNZ}(t) \qquad (6)$$

Next, we analyze in detail the individual contribution from each of these singularities.

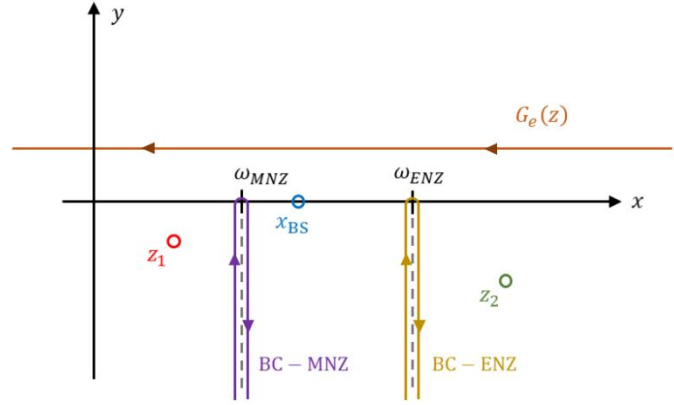

*Fig. 2. Sketch of the complex plane including the integration contour for the resolvent $G_e(z)$ and its singularities. The unstable poles, bound states (poles in the real axis) and branch cuts are indicated.*

<u>Contribution from the poles:</u> The poles of the resolvent are found at positions in the complex plane $z_\beta = x_\beta + i y_\beta$ corresponding to the solutions to the implicit equation $z_\beta = \omega_0 + \Sigma_e(z_\beta)$. Each contributes to the probability amplitude as an exponential decay term, $R_\beta e^{-i z_\beta t}$, where the initial-time ($t=0$) contribution can be computed via the residue theorem, which for a pole of order one can be compactly written as $R_\beta = \left(1 - [\partial_z \Sigma_e(z)]_{z=z_\beta}\right)^{-1}$.

Figure 2 schematically depicts the position of the three different poles that are associated with this waveguide. Figs. 3a and 3b show the projections of the poles on the real $x_\beta$ and imaginary $y_\beta$ axis, respectively, as functions of the transition frequency of the emitter $\omega_0$. The existence of poles critically depends on the position of $\omega_0$ with respect to the dispersion properties of the waveguide. Specifically, the three different poles are: (i) an unstable pole $z_1$, i.e., a pole with $y_\beta < 0$, that exists when the transition frequency of the emitter lies within the low-frequency (negative index) propagating band, $\omega_0 \leq \omega_{MNZ}$; (ii) an unstable pole $z_2$ that exists when the transition frequency of the emitter lies within the high-frequency (positive index) propagating band, $\omega_0 \geq \omega_{ENZ}$; and (iii) a bound state $z_{BS} = x_{BS}$, i.e., a pole exactly located on the real axis, $y_{BS} = 0$, which exists for $\omega_{BS} \geq \omega_{MNZ}$.

As a consequence of the asymmetric edges of the band gap, the existence of the bound state is not guaranteed for all values of $\omega_0$. Specifically, the diverging and negative values of $\Delta\omega(\omega)$ near $\omega_{ENZ}$ ensure that a real solution to $x_{BS} = \omega_0 + \Delta\omega(x_{BS})$ exists even for very large values of $\omega_0$, with $x_{BS} \sim \omega_{ENZ}$ (see Fig. 3a). This behavior has been observed at the band-edges of other photonic nanostructures[38–41]. However, because $\Delta\omega(\omega)$ approaches zero at $\omega_{MNZ}$, there are no real solutions to $x_{BS} = \omega_0 + \Delta\omega(x_{BS})$ for $\omega_0 < \omega_{MNZ}$. For the same reason, there is a soft transition where an unstable pole is smoothly transformed into a bound state as the emitter's frequency decreases below $\omega_{MNZ}$. This property might be of interest for shifting between dissipative and coherent interactions by tuning the frequency of the emitter.

The projections of the poles onto the imaginary axis indicate how fast the contributions from the poles decay. In general, the decay rate, i.e., the imaginary part of the projection of the self-energy on the real axis, $\Gamma(x_\beta)$, provides a good approximation of the behavior of the poles (see Fig. 3c). However, the decay rate $\Gamma(x_\beta)$ is predicted to diverge at the ENZ frequency. This is an expected, unrealistic behavior. As shown in Fig. 3c., the nonperturbative theory predicts that the imaginary part of the pole saturates to a finite value in the $\omega_0 \to \omega_{ENZ}$ limit, which in turn imposes a limit on the speed on the decay process. In practice, a divergent decay rate will never be observed due to dissipation losses and fabrication tolerances. However, this result highlights that nonperturbative effects act as an additional limiting factor on the enhancement of the decay rate in dispersive waveguides.

Fig. 4(a) depicts the magnitude of initial-time ($t = 0$) contributions of the poles as a function of the transition frequency of the emitter $\omega_0$. As expected, once the emitter is located well within the propagating band, the response of the system is dominated by the contribution of a single unstable pole. However, a more complex scenario takes place near the band edges. When the emitter's frequency equals the ENZ frequency, $\omega_0 = \omega_{ENZ}$, the contribution of the bound state and the higher-band unstable pole are of equal significance. This situation leads to interference effects in the decay dynamics. On the other hand, when the emitter's frequency equals the MNZ frequency, $\omega_0 = \omega_{MNZ}$, the contributions from the poles suddenly drop to zero; and the dynamics become entirely dominated by the contribution from the branch cuts. This behavior is justified by the fact that the derivative of the self-energy diverges at $z = \omega_{MNZ}$. It is clear that it is crucial to account for nonperturbative phenomena in both cases to properly access the exact decay dynamics near the band edges.

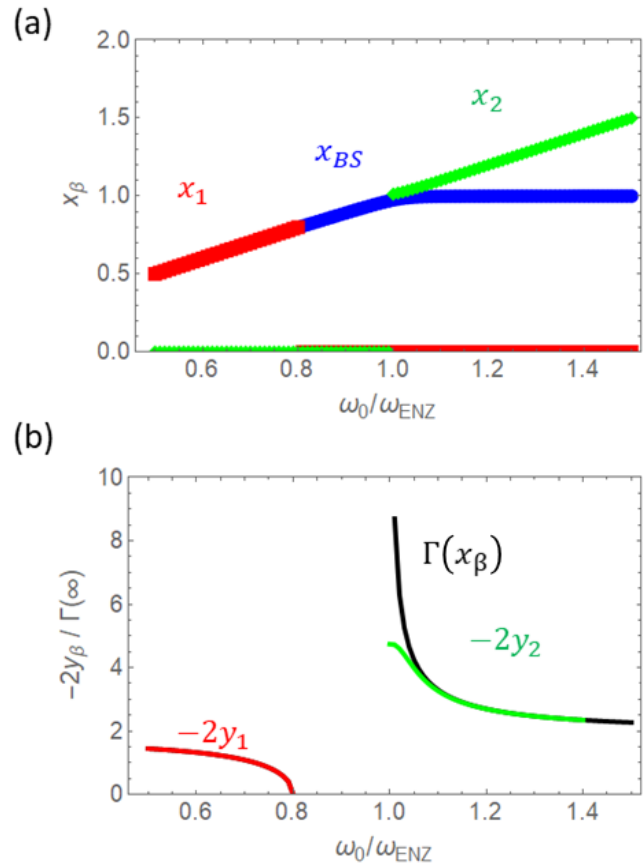

Fig. 3. Locations of the poles, $z_\beta = x_\beta + iy_\beta$, in the complex plane including two unstable poles $\beta = 1,2$ and a bound state $\beta = BS$, as described by their projections onto the (a) real $x_\beta$ and

*(b) imaginary $y_\beta$ axis. The evaluation of the decay rate $\Gamma(x_\beta)$ is included for the sake of comparison.*

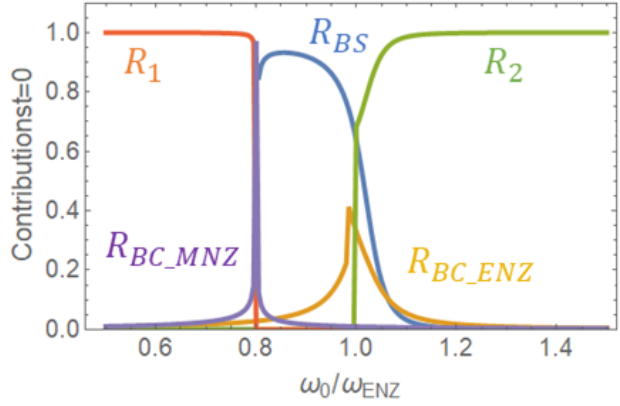

*Fig. 4. Magnitude of the initial-time $(t = 0)$ contributions for the probability amplitude, including the unstable poles associated with the low-frequency $R_1$ and high-frequency $R_2$ propagating bands, the bound state $R_{BS}$, and the mu-near-zero $R_{BC\_MNZ}$ and epsilon-near-zero $R_{BC\_ENZ}$ branch cuts.*

<u>Contributions from the branch cuts:</u> The contributions from the branch cuts, $C_{BC\_ENZ}(t)$ and $C_{BC\_MNZ}(t)$, are harder to analyze as they require numerical integration along vertical paths in the complex plane (see Fig. 2). Intuitively, one can understand these contributions as arising from a collection of singularity contributions leading to a fractional decay. This intuition is ratified by observing the long-time limits of the integral which yield fractional decay rates: $\lim_{t\to\infty} C_{BC\_ENZ}(t) \propto t^{-\frac{3}{2}}$ and $\lim_{t\to\infty} C_{BC\_MNZ}(t) \propto t^{-\frac{1}{2}}$. Again, we find that the asymmetric band-edges present very different properties. Specifically, the ENZ branch cut contributes with a $t^{-\frac{3}{2}}$ power law similar to that predicted for other slow-light waveguides[30]. On the other hand, the MNZ branch cut exhibits a slower power law decay, $t^{-\frac{1}{2}}$. The initial-time contributions, $C_{BC\_ENZ}(t = 0)$ and $C_{BC\_MNZ}(t = 0)$, depicted in Fig. 4(a) illustrate that the branch-cuts only produce a significant contribution when the transition frequencies of the emitter are near the branch points given by the ENZ and MNZ frequencies.

<u>Examples of decay dynamic profiles:</u> To finalize, Fig. 5 displays some representative examples of the time evolution of the survival probability of the excited state, $|C_e(t)|^2$, as the emitter's frequency is scanned through the dispersion profile of the waveguide. These examples serve to highlight the variety of temporal profiles that can take place when different singularities in the complex plane become dominant.

First, when the emitter is tuned to the low-frequency propagating band, e.g., $\omega_0 = 0.6\,\omega_{ENZ}$, the decay dynamics exhibit a near exponential decay. This property is justified by the fact that the response is dominated by a single unstable pole. Second, when the emitter is tuned to the MNZ frequency: $\omega_0 = 0.8\,\omega_{ENZ} = \omega_{MNZ}$, the time-evolution displays the slow fractional decay that arises from the branch-cut contribution. Therefore, although the MNZ band-edge enables a smooth transition from an unstable pole to a bound state, the decay dynamics will exhibit a residual and slow fractional decay when the emitter is tuned very close to that band-edge. When the emitter is tuned to the middle of the band gap, e.g., $\omega_0 = 0.85\,\omega_{ENZ}$, the temporal profile is characterized by a fast decay into a long-lived bound state. Interestingly, when the emitter is tuned to the ENZ frequency: $\omega_0 = \omega_{ENZ}$, the decay dynamics exhibit a complex and slow oscillatory decay. Although a fast decay would be expected from the weak coupling regime because of the Purcell enhancement at the band-edge, the interference between the unstable pole, bounds state and branch cut contributions result in a significantly different profile. Finally, when the emitter is tuned well into the high-frequency propagating band, e.g., $\omega_0 = 1.2\,\omega_{ENZ}$, the decay is again dominated by a near-exponential profile associated with a single unstable pole.

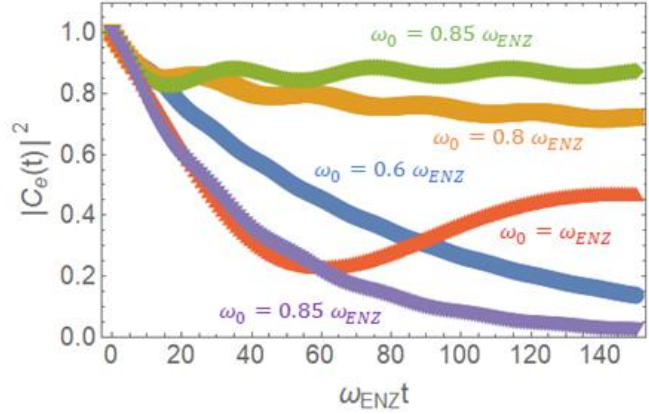

*Fig. 5. Representative examples of the time evolution of the survival probability of the excited state, $|C_e(t)|^2$, for different transitions frequencies. The case studies include the emitter being tuned to (i) the low-frequency propagating band at $\omega_0 = 0.6\,\omega_{ENZ}$, (ii) the mu-near-zero frequency at $\omega_0 = 0.8\omega_{ENZ} = \omega_{MNZ}$, (iii) the middle of the band gap at $\omega_0 = 0.85\,\omega_{ENZ}$, (iv) the epsilon-near-zero frequency at $\omega_0 = 0.85\,\omega_{ENZ}$, and (v) the high-frequency propagating band at $\omega_0 = 1.2\omega_{ENZ}$.*

In conclusion, our results illustrate the wealth in decay dynamics phenomena that can be observed in metamaterial waveguides that have a complex dispersion profile. We have shown that not all slow-light band-edges are equal. Some present a divergent behavior of the decay rate that is of particular interest for bright and deterministic photon sources, while others enable the smooth transition from an unstable pole to a bound state that is of particular interest for entanglement generation and many-body physics. Finally, it is crucial to properly account for the branch cut contributions since they lead to fractional decay rates whose properties also depend on the characteristics of the band-edge. We expect that our nonperturbative decay dynamics theory will be very relevant for a number of metamaterial waveguides in the optical regime, as well as for microwave superconducting circuits with similar dispersion profiles.

## References

I.L. acknowledges support from H2020-ERC-2020-StG-94850, Ramón y Cajal fellowship RYC2018-024123-I and project RTI2018-093714-J-I00 sponsored by MCIU/AEI/FEDER/UE.